# Perfect conversion of a TM surface-wave into a TM leaky-wave by an isotropic periodic meta-surface printed on a grounded dielectric slab

Svetlana N. Tcvetkova, Enrica Martini, Sergei A. Tretyakov, and Stefano Maci

*Abstract*— **This paper presents an exact solution for a perfect conversion of a TM-polarized surface wave (SW) into a TM-polarized leaky-wave (LW) using a reciprocal and lossless penetrable metasurface (MTS) characterized by a scalar sheet impedance, located on a grounded slab. In contrast to known realizations of leaky-wave antennas, the optimal surface reactance modulation which is found here ensures the absence of evanescent higher-order modes of the field Floquet-wave expansion near the radiating surface. Thus, all the energy carried by the surface wave is used for launching the single inhomogeneous plane wave into space without accumulation of reactive energy in the higher-order modes. It is shown that the resulting penetrable MTS exhibits variation from an inductive to a capacitive reactance passing through a resonance. The present formulation complements a previous paper of the authors in which a perfect conversion from TM-polarized SW to TE-polarized LW was found for impenetrable boundary conditions. The solution here takes into account the grounded slab dispersion and it is convenient for practical implementation.**

*Index Terms*— **Metasurface, Surface Waves, Floquet Waves, Leaky Waves, Antennas.**

## I. INTRODUCTION

Surface-to-leaky wave conversion is one of the classical problems in antennas, plasmonics, and nanophotonics [1]-[15]. In recent antenna applications [16]-[22] this conversion is obtained by the interaction of a cylindrical SW excited by a point source, with a curvilinear type modulation obtained by printing subwavelength patches on a grounded plane. In particular, in [23], the local interaction is studied by assuming a local plane SW wavefront interacting with a local boundary value problem (BVP) with unidirectional periodic modulation. This periodic BVP can be formulated by an infinite Floquet mode expansion [23] using an extension of the Oliner-Hessel method for impenetrable (Leontovitch-type) boundary condition [5] which accounts for the slab Green's function. Therefore, the periodic BVP assumes strategic importance in the design of a very flexible class of flat antennas. In any of these periodic BVP solutions, the vicinity of the reactive boundary contains fields of an infinite number of higher-order Floquet harmonics, which store reactive energy, producing a dispersive effect that limits the antenna bandwidth. This happens even if the penetrable MTS on top of the slab exhibits a gentle periodic variation of parameters, for instance of a sinusoidal form. Therefore, there is interest in finding a particular functional form of the modulating MTS reactance that allows for limiting as much as possible storage of reactive energy close to the surface, still having a "perfect" SW-to-LW conversion where only the 0 and -1 modes are present in the exact expansion.

Recently, new methods have been proposed for efficient conversion of a homogeneous plane wave into a surface wave using metasurfaces [9], [24]. In [16], a metasurface system able to route space waves via surface waves has been introduced. The system is synthesized based on a momentum transfer approach using phase-gradient metasurfaces and tested experimentally.

In [24], the problem of perfect conversion has been faced by using an impenetrable impedance model. Similarly to that work, the problem of the perfect conversion for the case of a single harmonic of a surface wave into a single harmonic of an inhomogeneous plane wave was considered in [25]. Despite this model does not describe the slab dispersion properly, it allows to find a simple analytical closed-form solution which is useful as a guideline. However, results for the point-wise lossless implementation can be obtained only for TM-SW to TE-LW conversion with an anisotropic metasurface. In [26], the authors describe a solution with only one propagating plane wave radiating from a partially transparent wall of a waveguide realized using a bianisotropic metasurface. However, the structure proposed for realization is rather complicated and bulky compared to conventional leaky-wave antennas used in practice.

In this paper, we introduce a theoretically perfect wave converter based on a single-sheet penetrable metasurface, which can be practically realized in a simple way as array of metal patches or slots in a metallic plate over a grounded dielectric slab. We have found that using an appropriate functional form of the isotropic metasurface leads to a perfect TM-TM

This work was supported in part by the Academy of Finland (project 287894), Nokia Foundation (project 201910452), HPY Research Foundation and in part by the University of Siena.

S. N. Tcvetkova and S. A. Tretyakov are with the Department of Electronics and Nanoengineering, Aalto University, P.O. 15500, FI-00076 Aalto, Finland (e-mail: svetlana.tcvetkova@aalto.fi).
S. Maci and E. Martini are with Department of Information Engineering and Mathematics,
University of Siena, Via Roma 56, 53100, Siena, Italy (e-mail: macis@dii.unisi.it)

conversion solution. This solution is presented in the next section.

## II. PROBLEM GEOMETRY AND PERIODIC BVP

Let us consider an MTS characterized by a periodic lossless impedance (Fig. 1). The MTS can be constituted by lossless periodically arranged subwavelength patches printed at $z = 0$ level over a lossless grounded dielectric slab of relative permittivity $\varepsilon_r$ and thickness $h$ (Fig. 1a). However, for the purpose of this paper, the MTS will be considered as a homogenized surface with $x$-axis periodicity dependence (Fig. 1b). In the following, $k_0 = \omega\sqrt{\varepsilon_0\mu_0}$ and $\zeta = \sqrt{\mu_0/\varepsilon_0}$ denote the free-space wavenumber and impedance, respectively.

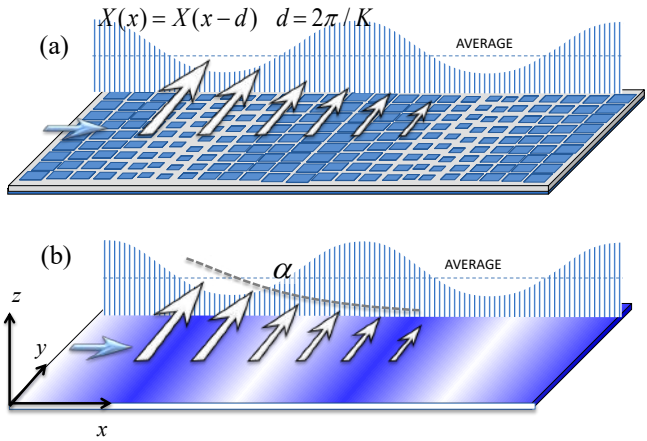

Fig. 1 (a) Scalar printed elements on a dielectric slab and (b) its homogenized version.

### A. Isotropic boundary condition

The MTS can be modeled by isotropic *continuous* impedance boundary conditions (BCs). If the metallic-patch layer has infinitesimal thickness, one may assume $E_x|_{0^+} = E_x|_{0^-} = E_x$. Assuming and suppressing time dependence $\exp(j\omega t)$, the MTS is modeled by using "transparent" isotropic BCs, defined by [23]

$$E_x = jX\left(H_y|_{0^+} - H_y|_{0^-}\right) = jX\,J_x,\tag{1}$$

where $X$ is the surface reactance and $J_x$ is the average electric surface current density flowing in the reactive sheet. In the absence of losses, $X$ is a real number. The model of transparent reactance in (1) is accurate when applied to a general wave field since it is almost independent of the $x$-component of the complex wavenumber $k_x$ in a quite large frequency range [27].

### B. TM- SW in the absence of modulation

In the presence of a uniform (i.e. non-modulated) impedance of value $X = X_0$, the structure supports a SW with currents

$J_{x0} = J_0 e^{-j\beta_{sw}x}$. The dispersive wavenumber $\beta_{sw}$ is the TM solution of the local transverse resonance equation over $X_0$ given by

$$\left[\frac{1}{X^+(\beta_{sw})} + \frac{1}{X^-(\beta_{sw})} + \frac{1}{X_0}\right] = 0,\tag{2}$$

where

$$X^+(k_x) = -\zeta\frac{\sqrt{k_x^2 - k_0^2}}{k_0},\tag{3}$$

$$X^-(k_x) = \zeta\frac{\sqrt{\varepsilon_r k_0^2 - k_x^2}}{k_0\varepsilon_r}\tan\left(h\sqrt{\varepsilon_r k_0^2 - k_x^2}\right)\tag{4}$$

are the TM reactances of a $z$-directed transmission line toward free space and toward the ground, respectively (see Figure 2). We note that in (2) the functions in (3) and (4) are evaluated for $k_x = \beta_{sw}$. In the following sections, the same functions will be evaluated at different complex wavenumbers.

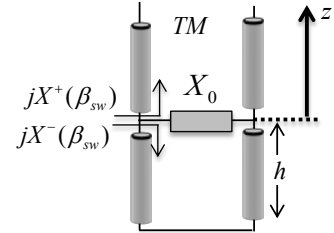

Fig. 2. $z$-directed transmission-line network for finding surface-wave dispersion for a uniform penetrable impedance of value $X_0$.

## III. REACTANCE SYNTHESIS FOR TWO MODES ONLY

In the presence of a periodic modulation of the reactance $X(x) = X(x - d)$ for $x > d$, where $d$ is the period of modulation, the energy transported by the SW leaks from the surface, transforming the bounded SW into a LW. The dominant component of the currents is substantially determined by the average transparent reactance and the global currents can be thought of as a result of interference between the SW over $X_0$ and the oscillation of the reactance. The 0-indexed mode of the Floquet expansion is given by

$$J^{(0)} = J_0 e^{-jk^{(0)}x}.\tag{5}$$

We denote $J^{(0)}$ as "0-mode" current, where $J_0$ is its constant amplitude, $k_x = k_x^{(0)} = \beta_x^{(0)} - j\alpha_x$ is the complex wavenumber of the surface wave (0-mode), and $x$ is the axis along the surface. The attenuation constant $\alpha_x$ is associated with the transfer of energy from the 0-mode to the radiating mode during the propagation path.

We would like to synthesize a periodic reactance $X$ that allows for the presence of only two modal currents in the radial Floquet Wave (FW) expansion, that is the 0-mode and the -1-mode (leaky mode). Defining $K = 2\pi/d$, we therefore assume that the exact current distribution is given by

$$J_x = J^{(0)} + J^{(-1)} \tag{6}$$

with $J^{(0)}$ defined in (5) and

$$J^{(-1)} = J_{-1} e^{-j(k^{(0)}-K)x} \tag{7}$$

where $J_{-1}$ is a constant amplitudes of the the -1 indexed mode, and $k_x = k_x^{(-1)} = k_x^{(0)} - K$ is the complex wavenumber of the leaky wave (-1-mode).

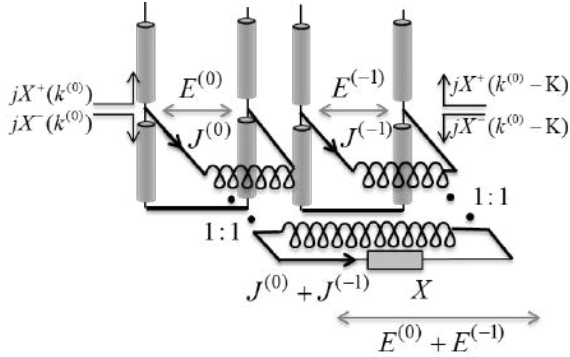

Fig. 3 Equivalent circuits for coupling the two $z$-directed transmission lines associated with the (0) and (-1) TM modes.

The FW-electric field mode is locally defined through the spectral Green's function (GF) of the grounded slab evaluated at the wavevector $k_x^{(0)}$ and $k_x^{(-1)}$ (see Fig. 3); i.e.,

$$E_x = E^{(0)} + E^{(-1)} \tag{8}$$

$$E^{(0)} = jX_{GF}^{(0)} \cdot J^{(0)} \quad ; \quad E^{(-1)} = jX_{GF}^{(-1)} \cdot J^{(-1)} \tag{9}$$

$$X_{GF}^{(0)} = -\frac{X^+\left(k_x^{(0)}\right)X^-\left(k_x^{(0)}\right)}{X^+\left(k_x^{(0)}\right) + X^-\left(k_x^{(0)}\right)} \tag{10}$$

$$X_{GF}^{(-1)} = -\frac{X^+\left(k_x^{(-1)}\right)X^-\left(k_x^{(-1)}\right)}{X^+\left(k_x^{(-1)}\right) + X^-\left(k_x^{(-1)}\right)} \tag{11}$$

in which $X^{\pm}\left(k_x\right)$ are defined in (3)-(4). These reactances are parallel to the sheet reactance and the input reactance of the grounded slab.

The exact solution of the BVP with two waves only is found by introducing (6) and (8) in (1), thus obtaining

$$X(J^{(0)} + J^{(-1)}) = X_{GF}^{(0)} \cdot J^{(0)} + X_{GF}^{(-1)} \cdot J^{(0)} \tag{12}$$

which is the transverse resonance equation for the problem. Figure 3 helps the interpretation of (12). Because of the absence

of forced excitation, the net active power balance is zero, therefore

$$\tfrac{1}{2}\mathrm{Re}\left[ jX(J^{(0)} + J^{(-1)})(J^{(0)} + J^{(-1)})^* \right] = 0 \tag{13}$$

where the left-hand side is the time average power totally furnished by the currents; this power should be zero since the power goes from the surface wave to the leaky wave in any spatial period, and there are no other modes in the expansion. Equation (13) provides an implicit evidence that $X$ is real on the dispersion curve in the absence of losses.

A two-wave solution exists if we find a solution of (12) with a real-valued reactance $X$. To this end we substitute in (12) the explicit values from (7) and simplify by dividing both sides by $e^{-jk^{(0)}x}$, thus obtaining

$$X(J_0 + J_{-1}e^{jKx}) = X_{GF}^{(0)}J_0 + X_{GF}^{(-1)} \cdot J_{-1}e^{jKx} \tag{14}$$

This equation can be rearranged as

$$X = \frac{X_{GF}^{(0)}e^{-jKx/2} + X_{GF}^{(-1)}\frac{J_{-1}}{J_0}e^{jKx/2}}{e^{-jKx/2} + \frac{J_{-1}}{J_0}e^{jKx/2}} \tag{15}$$

It may be seen that $X$ assumes real values if

$$J_{-1} = J_0 e^{-j\gamma} \tag{16}$$

$$X_{GF}^{(-1)} = X_{GF}^{(0)*} \tag{17}$$

where $\gamma$ is a constant phase. Indeed, denoting $X_{GF}^{(0)} = a + jb$ (15) leads to

$$X = \left( \frac{(a+jb)e^{-j(Kx/2 - \gamma/2)} + (a - jb)e^{j(Kx/2 - \gamma/2)}}{2\cos(Kx/2 - \gamma/2)} \right) \tag{18}$$

and therefore, one obtains the real solution

$$X = a + b\tan(Kx/2 - \gamma/2) \tag{19}$$

Defining $Kx_0 = \gamma$, one has

$$X = \mathrm{Re}\, X_{GF}^{(0)} + \mathrm{Im}\, X_{GF}^{(0)}\tan\left(\frac{2\pi}{d}(x - x_0)/2\right) \tag{20}$$

where $\mathrm{Re}\, X_{GF}^{(0)} = X_0$ plays the role of the average value of $X$. We note that the reactance in (20) is defined up to an arbitrary shift $x_0$ of the reference system, and therefore we can limit the investigation to $x_0 = 0$. The expression (20) depends on the complex parameter $k_x^{(0)}$. Therefore, the solution of the perfect TM-TM conversion exists if one finds $k_x^{(0)}$ such that $X_{\mathrm{GF}}^{(-1)} = X_{\mathrm{GF}}^{(0)*}$. The physical meaning of this condition, together with (16), is substantiated in Section V.

The exact solution of the two-wave problem is therefore determined if one finds a complex value of $k_x^{(0)}$ that respects the condition $X_{\mathrm{GF}}^{(-1)} = X_{\mathrm{GF}}^{(0)*}$. This condition in explicit form reads

$$\frac{1}{X^+\left(k_x^{(0)}\right)} + \frac{1}{X^-\left(k_x^{(0)}\right)} = \left[\frac{1}{X^+\left(k_x^{(0)}-K\right)} + \frac{1}{X^-\left(k_x^{(0)}-K\right)}\right]^* \quad (21)$$

This can be rewritten as

$$f\left(k_x^{(0)}\right) = f^*\left(k_x^{(0)} - K\right) \quad (22)$$

where $f\left(k_x\right) = \zeta \,/\, X^+\left(k_x\right) + \zeta \,/\, X^-\left(k_x\right)$, that is

$$f\left(k_x\right) = \frac{\varepsilon_r \cot\left(k_0 h \sqrt{\varepsilon_r - \left(k_x/k_0\right)^2}\right)}{\sqrt{\varepsilon_r - \left(k_x/k_0\right)^2}} - \frac{1}{j\sqrt{1 - \left(k_x/k_0\right)^2}} \quad (23)$$

The solution for a desired pointing angle of $\theta$ can be found as nulls of a real positive function, i.e.,

$$\left| f\left(\beta_x^{(0)} - j\alpha_x\right) - f^*\left(k_0 \sin\theta - j\alpha_x\right)\right| = 0 \quad (24)$$

where $\beta_x^{(0)} - K = k_0 \sin\theta$. The branch of the square root in (23) is taken in the appropriate way for forward and backward LW, as described in the next section.

## IV. FORWARD AND BACKWARD LW

The local dispersion equation of an inhomogeneous plane wave in free space yields

$$\left(\beta_x^{(0)}\right)^2 - \alpha_x^2 + \left(\beta_z^{(0)}\right)^2 - \left(\alpha_z^{(0)}\right)^2 = k_0^2 \quad (25)$$

$$\left(\beta_x^{(0)} - K\right)^2 - \alpha_x^2 + \left(\beta_z^{(-1)}\right)^2 - \left(\alpha_z^{(-1)}\right)^2 = k_0^2 \quad (26)$$

where $\alpha_z^{(n)} = -\operatorname{Im} k_z^{(n)}$, $\beta_z^{(n)} = \operatorname{Re} k_z^{(n)}$ $(n = 0, -1)$, with

$$k_z^{(0)} = \sqrt{k_0^2 - \left(k_x^{(0)}\right)^2} \text{ and } k_z^{(-1)} = \sqrt{k_0^2 - \left(k_x^{(0)} - K\right)^2} \quad (27)$$

Eqs. (25)-(26) allow for univocally defining all the parameters of the SW and LW as functions of $k_x^{(0)}$. According to well-known concepts and terminology, when the signs of $\beta_x^{(0)}$ and $\beta_x^{(-1)} = \beta_x^{(0)} - K = k_0 \sin\theta$ are the same (i.e., the wavefront propagation has the same direction as the power flow), the LW is called "forward LW"(FLW), and when they are opposite, it is called "backward LW" (BLW). While the BLW is "proper", namely it attenuates toward positive $z$-axis $(\alpha_z^{(-1)} = -\operatorname{Im} k_z^{(-1)} > 0)$, the FLW is "improper" (namely $(\alpha_z^{(-1)} = -\operatorname{Im} k_z^{(-1)} < 0)$.

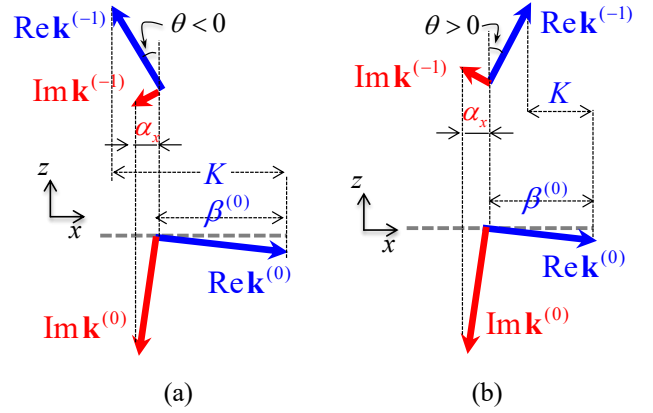

Fig. 4 Directions of the SW wavevector and (a) BLW (b) FLW.

To visualize the two cases, Figure 4 presents a sketch of the orientation of $\operatorname{Re}(\mathbf{k}^{(n)}) = \beta_x^{(n)}\mathbf{x} + \beta_z^{(n)}\mathbf{z}$, and $\operatorname{Im}(\mathbf{k}^{(n)}) = -\alpha_x^{(n)}\mathbf{x} - \alpha_z^{(n)}\mathbf{z}$ for the BLW (Fig. 3a) and FLW (Fig. 3b). Accordingly the branch of the imaginary part of the last term in (27) have to be as follows

$$\operatorname{Im}\sqrt{1 - \left(\sin\theta - j\alpha_x/k_0\right)^2} < 0 \quad (\theta \leq 0, \text{ BLW}) \quad (28)$$

$$\operatorname{Im}\sqrt{1 - \left(\sin\theta - j\alpha_x/k_0\right)^2} > 0 \quad (\theta > 0, \text{ FLW}) \quad (29)$$

The 0-mode is obviously proper, namely $\operatorname{Im}\sqrt{1 - \left(\beta_x^{(0)} - j\alpha_x\right)^2/k_0^2} < 0$.

The procedure for designing the reactance is summarized here for convenience. First, one fixes the value of the dielectric permittivity and thickness of the slab, the operating frequency and the pointing angle $\theta$ of the beam. Next, one finds the solution of (22) (or (24)) which provides $k_x^{(0)}$. Finally, the reactance is found from (20) with the use of (10) and (11) with the choice of the branch defined in (28) and (29).

## V. POWER BALANCE

Since (16) and (17) ensure $X$ to be a real value, (13) implies that the currents $J^{(0)}$ and $J^{(-1)}$ do not complexly produce active power, but only exchange power with each other in each period. In particular, the power density transported by the SW and LW contribution per unit surface are given by

$$P_{\mathrm{sw}} = -\frac{1}{2}\operatorname{Re}\left\{J^{(0)*} \cdot E^{(0)}\right\} = -\frac{1}{2}\operatorname{Im}\left\{X_{\mathrm{GF}}^{(0)}\left|J_0\right|^2\right\}$$ and

$$P_{\mathrm{lw}} = -\frac{1}{2}\operatorname{Re}\left\{J^{(-1)*} \cdot E^{(-1)}\right\} = -\frac{1}{2}\operatorname{Im}\left\{X_{\mathrm{GF}}^{(-1)}\left|J_{-1}\right|^2\right\},$$ respectively, and the reactive power stored per unit surface area is given by

$$W_{\text{sw}} = -\frac{1}{2}\text{Re}\left\{X_{\text{GF}}^{(0)}\left|J^{(0)}\right|^2\right\} \quad \text{and} \quad W_{\text{lw}} = -\text{Re}\left\{X_{\text{GF}}^{(-1)}\left|J^{(-1)}\right|^2\right\},$$

respectively. The conditions (16) and (17) are therefore equivalent to

$$P_{\text{sw}} + P_{\text{lw}} = 0 \qquad (30)$$

$$W_{\text{lw}} = W_{\text{sw}} \qquad (31)$$

Eq.(30) means that there is a perfect conversion of power from the surface to the leaky wave and (31) means that there is an equal amount of energy stored in the fields of the SW and in the LW fields.

## VI. DISPERSION EQUATION

The dispersion equation in (24) subjected to (28)-(29) can be solved numerically using conventional MATLAB routines. In the examples presented next the dielectric permittivity of the slab assumes the three values $\varepsilon_r = 3,6,15$. We observe that for a fixed $\varepsilon_r$ the solution $k_x^{(0)}/k_0$ depends only on $k_0 h$ and $k_0 d$. Figure 5 presents, for each permittivity, the real and imaginary part of the solution $k_x^{(0)}/k_0$ for two cases, denoted by continuous and dashed lines, relevant to broadside radiation and a backward LW.

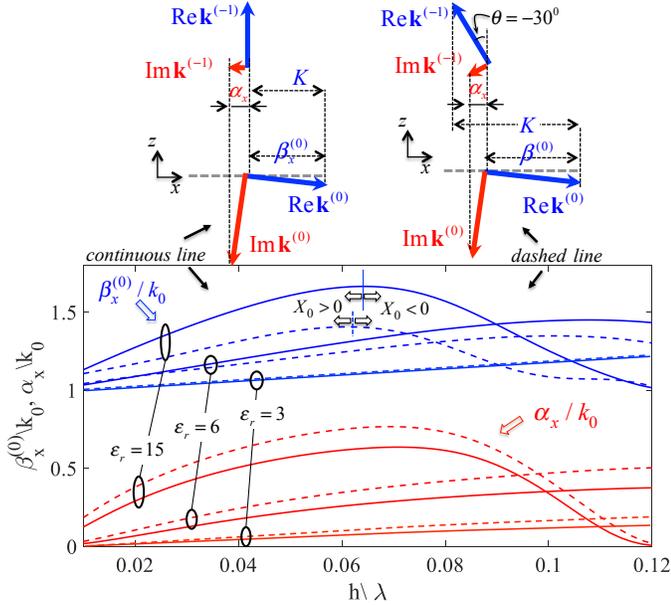

Fig. 5. Real and imaginary parts of $k_x^{(0)}/k_0$ as solutions of eq. (22) for various dielectric permittivities of the slab. Continuous line: broadside beam; dashed lines: backward beam at 30° counterclockwise from the normal. Insets illustrate the directions of the wavevectors of the two waves. The regions for which the average impedance $X_0 = \text{Re}\,X_{\text{GF}}^{(0)}$ is capacitive and inductive are shown for the case of the relative permittivity equal to 15.

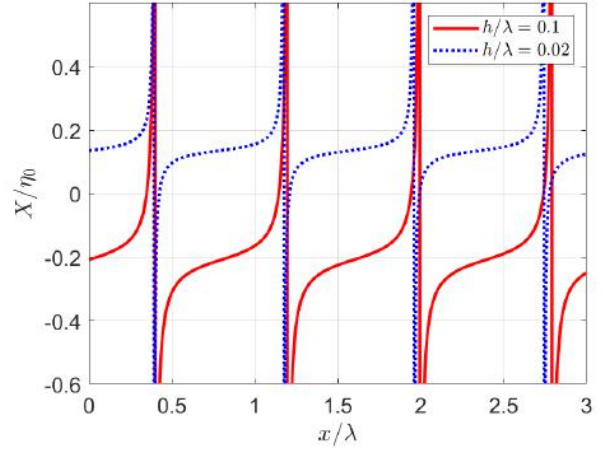

Fig. 6. Surface reactance at the frequency $f = 10$ GHz for the dielectric permittivity of the slab $\varepsilon_r = 15$, for the thicknesses of the slab $h/\lambda = 0.1$ (red line) and $h/\lambda = 0.02$ (blue dotted line) for the broadside beam case.

The solution is found for the propagating 0-mode toward positive $x$; the numerical process also allows for finding the solution $-k_x^{(0)}/k_0$ which corresponds to propagation toward negative $x$, not plotted in the figure. The broadside-beam solution is obtained by setting $K = \beta_x^{(0)}$, namely $\beta^{(-1)} = 0$ in (22), solving

$$\left|f\left(\beta_x^{(0)} + j\alpha_x\right) - f^*\left(j\alpha_x\right)\right| = 0 \qquad (32)$$

The backward LW solution, radiating at 30 degree direction (dashed lines), where the angle is counted counterclockwise from the normal to the surface, is obtained by setting $\beta^{(-1)} = -k_0 \sin(30°)$, namely $\beta^{(0)} - K = -k_0 \sin(30°)$. From (24), this is equivalent to finding solution of

$$\left|f\left(\beta_x^{(0)} + j\alpha_x\right) - f^*\left(-\frac{1}{2}k_0 + j\alpha_x\right)\right| = 0 \qquad (33)$$

with choice of the branch in (28). Blue and red curves in Fig. 5 denote $\text{Re}(k_x^{(0)}/k_0) = \beta_x^{(0)}/k_0$ and $-\text{Im}(k_x^{(0)}/k_0) = \alpha_x/k_0$, respectively, as functions of the thickness of the slab within the range $h/\lambda \in (0.02-0.12)$. Calculations have been carried out at 10 GHz; it is understood, however, that one can re-scale the solution for a different frequency. The period which allows for the specified direction of the beam is $d = 2\pi/\text{Re}(k_x^{(0)})$.

The results show that by changing the thickness of the slab one can change the leakage rate and, eventually, the pattern beamwidth and directivity of the antenna. It is interesting to note that, in the case $\varepsilon = 15$, for a fixed period corresponding to a certain desired radiation direction and for a desired decay rate $\alpha$, the surface can be realized by two different reactance profiles using different thicknesses $h$ and therefore different average impedance $X_0 = \text{Re}\,X_{\text{GF}}^{(0)}$. For instance, for the broadside case in Fig. 5, the same $d$ and $\alpha$ are found at $h/\lambda = 0.1$ and $h/\lambda = 0.02$ with two different periodic MTS reactances. For both cases in Fig. 5, the broadside one and the backward propagation at the angle $\theta = 30°$, we observe that for $h/\lambda < 0.06$ the average impedance is inductive and for $h/\lambda > 0.06$ it is capacitive.

Figure 6 shows the reactances as obtained from (20) (with $x_0 = 0$). In the two cases the average impedance Re $X_{GF}^{(0)} = X_0$ is negative (capacitive) or positive (inductive), respectively.

## VII. Considerations on Practical Implementation

In the exact solution (20), a tangent function appears which exhibits periodic singularities. These singularities can be approximated by printing elements which exhibit resonance of an equivalent $LC$ circuit of parallel type, for instance printed slots close to the resonance. As an example, we take the reactance solution of Fig. 6 for $h/\lambda = 0.02$ reported in Fig. 7 together with a layout of slots that emulate the required impedance. We note that in this case the poles and zeros of the solution are very close to each other. The slots are orthogonal to the propagation direction, and therefore the model is valid only for propagation along $x$. At the beginning of the period, the slots are very small and gradually become longer. When the slot becomes resonant, the impedance becomes infinite. At the end of the period, they become small again with a discontinuity. It is worth noting that implementing a solution with capacitive average values (e.g., the one in Fig. 6 for $h/\lambda = 0.1$) requires printed dipoles, which actually exhibit a series-type resonant behavior at the first resonance; therefore, printed dipoles should be used at the second resonance. Furthermore, the dipoles should be directed for this TM case in the direction of propagation; therefore, reproducing the impedance profile is critical. We are investigating a way to do it by using convoluted dipoles.

It is also worth noting that for the slot implementation, the approximation of homogeneous reactance is not a worry, while it could be significant how sensitive the solution is reproducing the desired profile close to the singularities. To this end, a sensitivity analysis to the deviation of the modulation profile wrt the ideal one is studied in Sec. VIIIB on the basis of the full-wave analysis carried out in Sec. VIIIA.

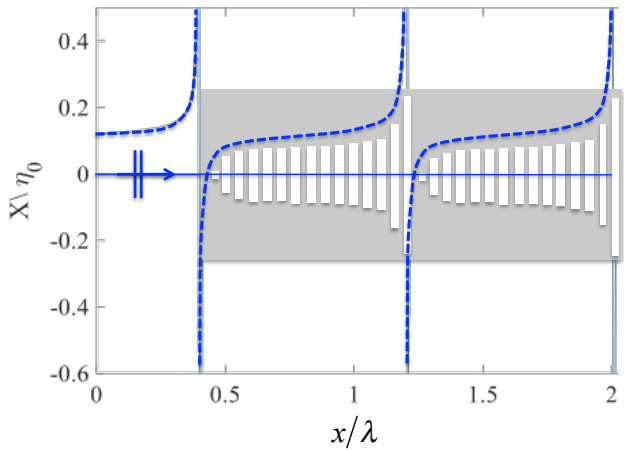

Fig. 7 Possible layout of a slot-type metasurface to reproduce the reactance shown in Fig. 6b. The layout is only indicative of a possible solution.

## VIII. Full-Wave Analysis and Validation

### A. Full-wave analysis of arbitrary periodic reactance

In order to test the solution and perform the sensitivity analysis we have developed a full-wave code for arbitrary homogenized penetrable reactance $X(x)$ of period $d$. The algorithm has been formulated by a periodic Method of Moments in which both currents and fields are represented through Floquet-wave (FW) basis functions. This analysis is a generalization to penetrable impedance and to arbitrary periodic profile of the Oliner method, which was presented in [5] for sinusoidal periodic Leontovich boundary conditions. In Appendix, it is shown that this generalization leads to a solution of the following linear algebraic infinite system of equations for the infinite number of unknowns $J_m$, which are the coefficients of the Floquet wave expansion of the currents that flow on the metasurface impedance:

$$\sum_m \chi_{n,m} J_m = 0 \qquad n=1,2...\infty \qquad (34)$$

$$\chi_{n,m} = X_{n-m} - X_{GF}\left(k_x^{(n)}\right)\delta_{n,m} \qquad (35)$$

where $k_x^{(n)} = k_x^{(0)} - 2\pi n/d$, $\delta_{n,m}$ is the Kronecker delta, and $X_p$ are the coefficients of the Fourier series expansion of the arbitrary periodic impedance. Eq. (34) should be truncated with a sufficiently high number of terms in order to reduce to a finite algebraic system, which admits solution only if $\det\left[\zeta_{n,m}\right] = 0$; the latter is the dispersion equation which identifies the complex value $k_x^{(0)}$ for the particular periodic reactance $X(x)$. Once the complex solution $k_x^{(0)}$ is found, one can derive the coefficients of the Floquet mode expansion as the entries of the eigenvector associated with the vanishing eigenvalue of the matrix $\left[\chi_{n,m}\right]$.

### B. Sensitivity analysis

The solution of (34) is used here to investigate the distribution of modal coefficients when the singularity of the tangent function profile is not sampled well by practical elements close to the singularity. The analysis has been performed for a slab of relative permittivity $\varepsilon_r = 15$, and $h/\lambda = 0.08$ and the leaky direction is broadside. Assuming a reactance of type $X(x) = a + b\tan\left(\pi x/d\right)$ and analyzing it through the full-wave code, we have found first that the solution of $\det\left[\chi_{n,m}\right] = 0$ is equivalent to imposing $X_{GF}(k_x^{(0)}) = X_{GF}^*(k_x^{(-1)}) = a + jb$, and then that the eigenvector of the vanishing eigenvalue has only two components ($n=0$, $n=-1$). The details of this demonstration are given in Appendix. The coefficients amplitude is shown in Fig. 8a, with the reactance profile in the inset. Fig. 8b, c, d show the results when $X(x) = a + b\tan\left(\pi x/d\right)$ is approximated using the Taylor expansion around $x=0$ of the tangent function, truncating the result in the period and repeating periodically the resulting function.

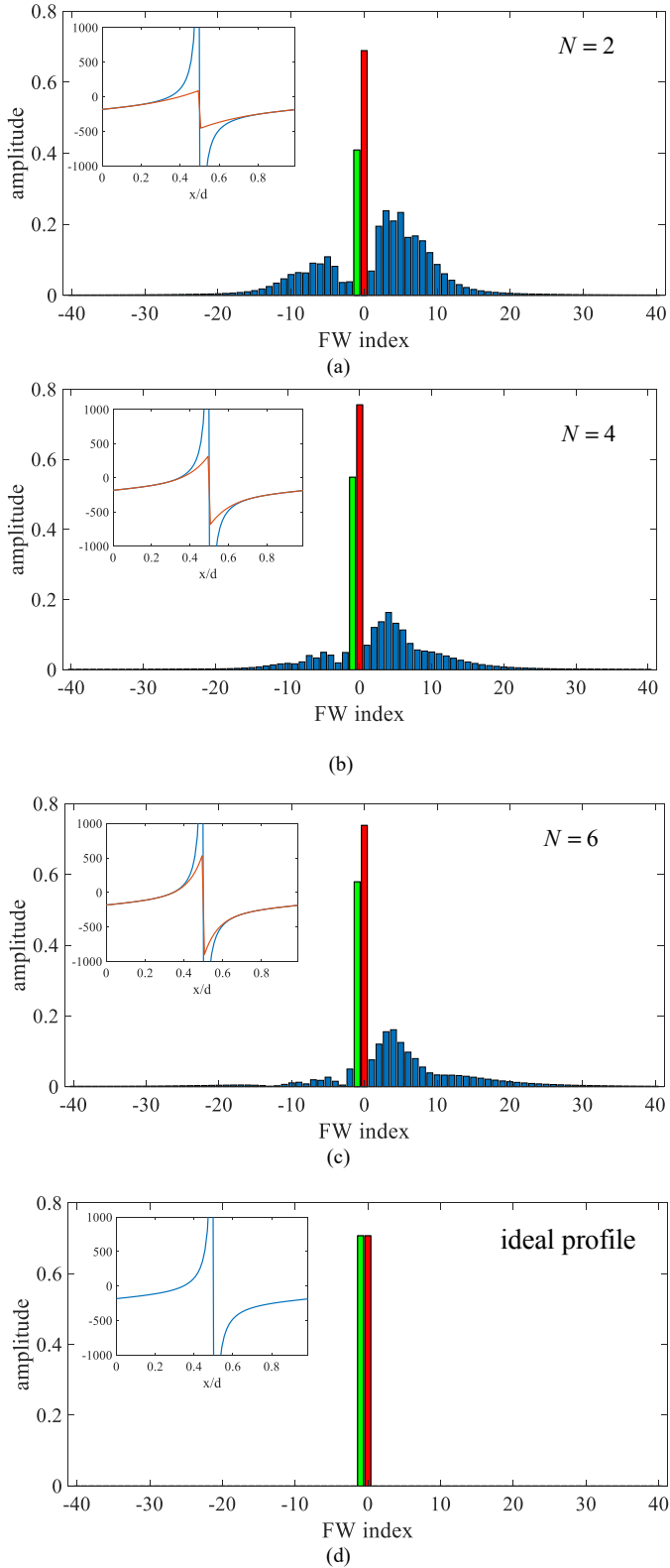

Fig. 8 Spectrum of FW amplitude of currents flowing in the metasurface when the reactance deviates from $X(x) = a + b\tan(\pi x / d)$ ($\varepsilon_r = 15$, $h / \lambda = 0.08$). The Taylor expansion of $X(x)$ is used around zero, truncated in one period, and periodically replicated. The results are obtained by the full-wave analysis in Section VIII A. (a), (b),(c) correspond to $N=2$, $N=4$, $N=6$ terms in the Taylor expansion, respectively The resulting approximation of the reactance is given in the inset in red; (d) presents the results for the exact representation of the tangent function.

The tangent function is approximated with $N$ terms of the Taylor expansion with $N=2,4,6$ (Figs. 8a,b,c) and next compared with the ideal profile (Fig. 8d) outcoming from the full wave analysis. This approximation emulates the case in which the resonance is not sampled properly by the elements. It is seen that decreasing the accuracy of the approximation close to the resonance, the amplitudes of the coefficients of the higher-order FWs increase and correspondingly the amplitude of the (-1) mode coefficient decreases wrt the amplitude of the (0) mode. This means that the energy *per unit cell* transferred from the 0 mode to the (-1) mode decreases.

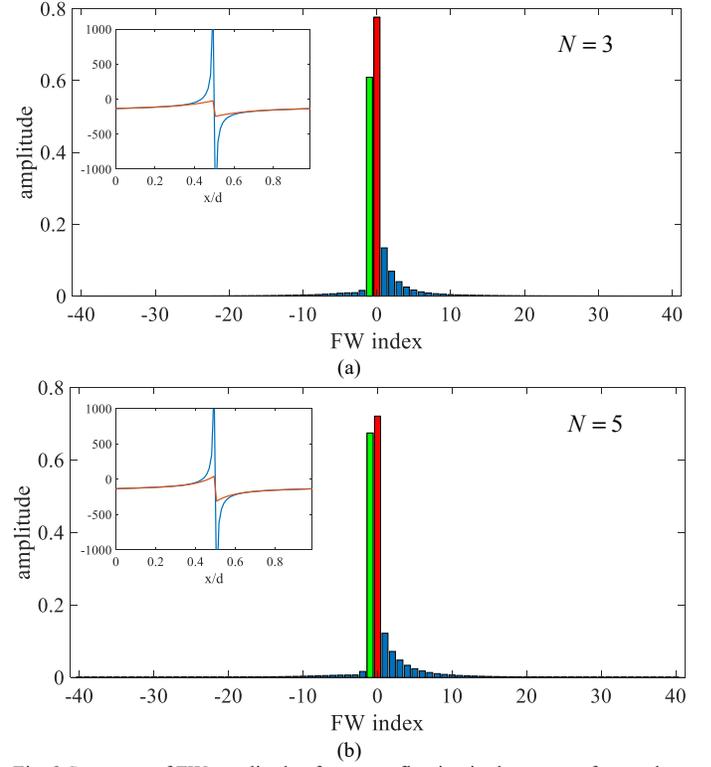

Fig. 9 Spectrum of FW amplitude of currents flowing in the metasurface. when the reactance deviates from $X(x) = a + b\tan(\pi x / d)$ ($\varepsilon_r = 3$, $h / \lambda = 0.08$). Only two cases of Taylor approximations are shown.

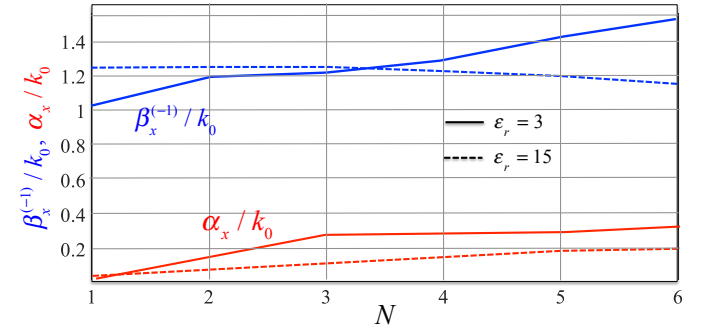

Fig. 10 Values of the real part and imaginary part of $k_x^{(-1)}$ for the cases show in Fig. 8 ($\varepsilon_r = 15$) and Fig. 9 ($\varepsilon_r = 3$). The value of the perfect case is $k_x^{(-1)} / k_0 = 1.48 - j0.6$ for $\varepsilon_r = 15$, and $k_x^{(-1)} / k_0 = 1.15 - j0.08$ for $\varepsilon_r = 3$.

The same exercise is repeated for $\varepsilon_r = 3$ and $h / \lambda = 0.08$ in Fig. 9 (the perfect approximation case is not presented since it gives the exact 2-mode approximation similar to the one in Fig. 8d).

The results show that the situation appears less critical with respect to the case with higher permittivity in Fig. 8. It should be observed that changing the approximation around the singularity also changes the value of the leaky wavenumber. Its behaviour is presented in Fig. 10. This means in practice that an inaccuracy in sampling the ideal profile of the metasurface impedance reactance not only impacts to the perfet modal conversion, but also changes the beam direction. Although this change can be compensated with a change of the periodicity, it implies less flexibility in the design.

## IX. CONCLUSIONS

We have presented a solution of a periodic isotropic transparent MTS impedance on a grounded dielectric slab, which corresponds to a perfect conversion from a TM-polarized surface wave to a TM-polarized leaky wave without higher-order Floquet harmonics in the exact expansion of the field solution. This solution has been found by imposing a reactive balance of the wave coupling in the absence of losses. The found impedance can be synthesized in order to ensure a certain beam direction and decay rate. Along the direction of propagation, the solution of transparent impedance exhibits alternance of positive and negative values around a certain average, passing through a resonance at each period. This suggests that a possible realization of this impedance can consist of a periodic distribution of printed subwavelength elements, where some are close to resonance, inside a distribution of non-resonant subwavelength printed elements.

It is also worth noting that the achievement of the two-mode only solution can have some drawbacks in terms of design flexibility. In fact, changing the leakage parameter to control the amplitude of the -1 mode on the same slab is not actually possible without simultaneously changing the period.

## APPENDIX

Let us assume to have a general periodic reactance $X(x)$ with the period $d$. Expanding both electric field and currents (1) in terms of Floquet modes, leads to

$$X(x)\sum_q J_q e^{-jk_{x,q}x} = \sum_p X_{GF}\left(k_x^{(p)}\right) J_p e^{-jk_{x,p}x} \qquad (36)$$

where $X_{GF}\left(k_x\right)$ is the Green function defined in (10), and $k_x^{(p)} = k_x^{(0)} - \dfrac{2\pi p}{d}$ are the Floquet-modes wavenumbers.

Expanding the reactance in Fourier series $X(x) = \sum_m X_m e^{-j\frac{2\pi m}{d}x}$ in (36) leads to

$$\sum_{m,q} J_q X_m e^{-j\frac{2\pi}{d}(m+q)x} = \sum_p X_{GF}\left(k_x^{(p)}\right) J_p e^{-j\frac{2\pi p}{d}x} \quad (37)$$

Here, the factor $k_x^{(p)} = k_x^{(0)} - 2\pi p / d$ has been deleted as present in both sides of (36). We note that both sides of (37) are periodic functions of period $d$. The left-hand side term can be therefore expanded in Fourier series by projecting on the term $\exp(j\frac{2\pi n}{d}x)$. This leads to Fourier expansion coefficient equal to $\sum_m J_m X_{n-m}$, namely the convolution between the coefficient expansion between reactance and currents. Since these coefficients must be equal to the ones at right hand side of (37), one obtains

$$\sum_m J_m X_{n-m} = X_{GF}\left(k_x^{(n)}\right) J_n \quad (38)$$

which is a system with an infinite number of equations and unknowns. The above can be easily rearranged as in (34)-(35). Eq. (38) can be also used to check the validity of the solution when $X(x) = a + b\tan(\pi x/d)$. In this case, the expression of the Fourier coefficients are $X_p = a\delta_{p,0} - jb(-1)^p \mathrm{sgn}(p)$ where $\delta_{p,0}$ is the Kronecker delta and sgn($p$)=1 for $p$ positive, -1 for $p$ negative, and 0 for $p$=0. We can specialize (38) for $n$ and $n$-1 and add the two equations, thus obtaining

$$\sum_m (X_{n-m} + X_{n-1-m}) J_m = X_{GF}\left(k_x^{(n)}\right) J_n + X_{GF}\left(k_x^{(n-1)}\right) J_{n-1} \quad (39)$$

With the defined values of $X_p$, all the terms of the summation exactly cancel each other except for the values $n = m$ and $n - 1 = m$, thus leading to

$$(a + jb)J_n + (a - jb)J_{n-1} = X_{GF}\left(k_x^{(n)}\right) J_n + X_{GF}\left(k_x^{(n-1)}\right) J_{n-1} \quad (40)$$

The latter can be rewritten as

$$\left[ X_{GF}\left(k_x^{(n)}\right) - (a+jb) \right] J_n + \left[ X_{GF}\left(k_x^{(n-1)}\right) - (a-jb) \right] J_{n-1} = 0 \quad (41)$$

which must be satisfied for any $n$. Specializing the latter expressions for $n$=0, one has

$$\left[ X_{GF}\left(k_x^{(0)}\right) - (a+jb) \right] J_0 + \left[ X_{GF}\left(k_x^{(-1)}\right) - (a-jb) \right] J_{-1} = 0 \quad (42)$$

which is identically satisfied when

$$X_{GF}\left(k_x^{(0)}\right) = X_{GF}^*\left(k_x^{(-1)}\right) = (a+jb) . \quad (43)$$

This result recovers the expression found in Section III. Furthermore, specializing (41) for $n=1$ and $n=-1$ and using (43) one gets $J_1 = 0$ and $J_{-2} = 0$, respectively. Proceeding iteratively for all $n$ in (41) $n \neq 0$ one finds that all the $I_n$ vanish for $n \neq 0$ and $n \neq -1$. This demonstrates that choosing the reactance like in (20) implies only two FWs in the solution for the electric currents.

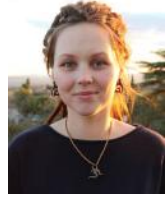

**Svetlana Tcvetkova** was born in Vyshny Volochyok, Russia, in 1991. She received M.Sc. in technical physics and M. Sc. (Tech) in computational engineering degrees in 2015 from Peter the Great St. Petersburg Polytechnic University, Russia and Lappeenranta University of Technology, Finland, respectively. In 2017, she was a visiting student researcher in the group of Prof. S. Maci at University of Siena, Italy. In 2019, she received the D. Sc. (Tech.) degree from Aalto University, Finland. She is currently working as a postdoctoral researcher in the group of Prof. S. Tretyakov at the Department of Electronics and Nanoengineering, Aalto University, Finland. Her main research interests include antennas, microwave measurements, and reflection/transmission properties of metasurfaces.

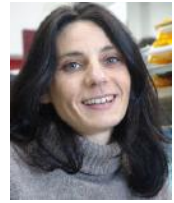

**Enrica Martini** (S'98-M'02-SM'13) was born in Spilimbergo (PN), Italy, in 1973. In 1998, she received the Laurea degree (cum laude) in telecommunication engineering from the University of Florence, Italy, where she worked under a one-year research grant from the Alenia Aerospazio Company, Rome, Italy, until 1999. In 2002, she received the PhD degree in informatics and telecommunications from the University of Florence and the Ph.D. degree in electronics from the University of Nice-Sophia Antipolis, under joint supervision.

In 2002, she was appointed Research Associate at the University of Siena, Italy. In 2005, she received the Hans Christian Ørsted Postdoctoral Fellowship from the Technical University of Denmark, Lyngby, Denmark, and she joined the Electromagnetic Systems Section of the Ørsted•DTU Department until 2007. From 2007 to 2017 she was a Postdoctoral Fellow at the University of Siena, Italy. From 2016 to 2018 she was the CEO of the start-up Wave Up Srl, Siena, Italy, that she co-founded in 2012. She is currently assistant professor at the University of Siena, Italy.

Dr. Martini was a co-recipient of the 2016 Schelkunoff Transactions Prize Paper Award and of the Best Paper Award in Antenna Design and Applications at the 11th European Conference on Antennas and Propagation.

Her research interests include metasurfaces, metamaterial characterization, electromagnetic scattering, antenna measurements, finite element methods and tropospheric propagation.

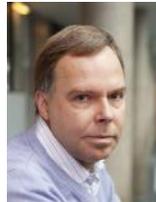

**Sergei A. Tretyakov** received the Dipl. Engineer-Physicist, the Candidate of Sciences (PhD), and the Doctor of Sciences degrees (all in radiophysics) from the St. Petersburg State Technical University (Russia), in 1980, 1987, and 1995, respectively. From 1980 to 2000 he was with the Radiophysics Department of the St. Petersburg State Technical University. Presently, he is professor of radio science at the Department of Electronics and Nanoengineering, Aalto University, Finland. His main scientific interests are electromagnetic field theory, complex media electromagnetics, metamaterials, and microwave engineering. He has authored or co-authored five research monographs and more than 300 journal papers. Prof. Tretyakov served as President of the Virtual Institute for Artificial Electromagnetic Materials and Metamaterials ("Metamorphose VI"), as General Chair, International Congress Series on Advanced Electromagnetic Materials in Microwaves and Optics (Metamaterials), from 2007 to 2013, and as Chairman of the St. Petersburg IEEE ED/MTT/AP Chapter from 1995 to 1998.

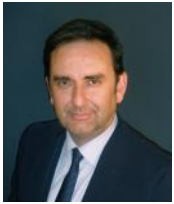

**Stefano MACI** (F04) received the Laurea Degree cum Laude at University of Florence in '87 and from '97 is a Professor at the University of Siena. His research interest includes high-frequency and beam representation methods, computational electromagnetics, large phased arrays, planar antennas, reflector antennas and feeds, metamaterials and metasurfaces. He was member the Technical Advisory Board of 12 international conferences, member of the Review Board of 6 International Journals, since 2000. He organized 25 special sessions in international conferences, and he held 10 short courses in the IEEE Antennas and Propagation Society (AP-S) Symposia about metamaterials, antennas and computational electromagnetics. In 2004-2007 he was WP leader of the Antenna Center of Excellence (ACE, FP6-EU) and in 2007-2010 he was International Coordinator of a 24-institution consortium of a Marie Curie Action (FP6). He has been Principal Investigator from 2010 of 6 cooperative projects financed by European Space Agency. In 2004 he was the founder of the European School of Antennas (ESoA), a post graduate school that presently comprises 34 courses on Antennas, Propagation, Electromagnetic Theory, and Computational Electromagnetics and 150 teachers coming from 15 countries. Since 2004 is the Director of ESoA.

Professor Maci has been a former member of the AdCom of IEEE Antennas and Propagation Society (AP-S), associate editor of AP-Transaction, Chair of the Award Committee of IEEE AP-S, and member of the Board of Directors of the European Association on Antennas and Propagation (EurAAP). From 2008 to 2015 he has been Director of the PhD program in Information Engineering and Mathematics of University of Siena, and from 2013 to 2015 he was member of the National Italian Committee for Qualification to Professor. He has been former member of the Antennas and Propagation Executive Board of the Institution of Engineering and Technology (IET, UK). He is the director of the consortium FORESEEN, presently involving 48 European Institutions, and principal investigator of the Future Emerging Technology project "Nanoarchitectronics" of the 8th EU Framework program. He was co-founder of 2 Spin-off Companies. He was a Distinguished Lecturer of the IEEE Antennas and Propagation Society (AP-S), and EuRAAP distinguished lecturer in the ambassador program. He was recipient of the EurAAP Award in 2014, of the IEEE Shelkunoff Transaction Prize 2015, of the Chen-To Tai Distinguished Educator award 2016, and of the URSI Dellinger Gold Medal in 2020. He is TPC Chair of the METAMATERIAL 2020 conference. In the last ten years he has been invited 25 times as key-note speaker in international conferences.

The research activity of Professor Maci is documented in 160 papers published in international journals, (among which 100 on IEEE journals), 10 book chapters, and about 400 papers in proceedings of international conferences. These papers have received around 7100 citations with h index 42.